\documentclass[aps,prl,showpacs,twocolumn,groupedaddress,superscriptaddress,floatfix]{revtex4}
\usepackage{amssymb}
\usepackage{graphicx}

\begin{document}

\title{Interferometry of Direct Photons in Central 
$^{208}$Pb+$^{208}$Pb Collisions at 158 A$\cdot$GeV.}

\author{ M.M.~Aggarwal}
 \affiliation{University of Panjab, Chandigarh 160014, India}
\author{ Z.~Ahammed}
 \affiliation{Variable Energy Cyclotron Centre, Calcutta 700064, India}
\author{ A.L.S.~Angelis$^*$}
 \affiliation{University of Geneva, CH-1211 Geneva 4, Switzerland}
\author{ V.~Antonenko}
 \affiliation{RRC ``Kurchatov Institute'', RU-123182 Moscow}
\author{ V.~Arefiev}
 \affiliation{Joint Institute for Nuclear Research, RU-141980 Dubna, Russia}
\author{ V.~Astakhov}
 \affiliation{Joint Institute for Nuclear Research, RU-141980 Dubna, Russia}
\author{ V.~Avdeitchikov}
 \affiliation{Joint Institute for Nuclear Research, RU-141980 Dubna, Russia}
\author{ T.C.~Awes}
 \affiliation{Oak Ridge National Laboratory, Oak Ridge, Tennessee 37831-6372, USA}
\author{ P.V.K.S.~Baba}
 \affiliation{University of Jammu, Jammu 180001, India}
\author{ S.K.~Badyal}
 \affiliation{University of Jammu, Jammu 180001, India}
% ...

\collaboration{WA98 collaboration}
\noaffiliation

\date{\today}

\begin{abstract}
  Two-particle correlations of direct photons were measured in central
  $^{208}$Pb+$^{208}$Pb 
collisions at 158 AGeV. The invariant interferometric radii were
  extracted for $100<K_T<300$ MeV/$c$ and compared to radii extracted 
from charged pion correlations.  The yield of soft direct photons,
  $K_T<300$ MeV/$c$, was extracted from the correlation strength 
and compared to theoretical  calculations.
\end{abstract}


\maketitle

Hanbury Brown-Twiss (HBT) interferometry provides a powerful tool
to explore the space-time dimensions of the emitting source 
created in elementary
particle or heavy ion collisions. Historically, such measurements have
concentrated on pion pair correlations, but have also been
applied to kaons, protons, and even heavy fragments \cite{QM-rew}.
Hadron correlations reflect the space-time extent of the 
emitting source at the time of freeze-out. 

The importance of direct photon Bose-Einstein interferometry for investigation of the 
history of heavy ion collisions, especially of the very early phase, 
has been extensively discussed in the literature
\cite{Srivastava,Timerman,Ornik,Peressounko}. 
It has been shown that the correlations of direct photons
in different ranges of
the photon average transverse momenta reflect different stages of the
collision. 
Therefore, photon-photon correlations can provide important 
information complementary to that obtained from hadron correlations. 
Unfortunately, photon interferometry is faced with considerable
difficulties compared to hadron interferometry due to
the small yield of photons emitted directly from the hot zone in 
comparison to the huge
background of photons produced by the electromagnetic decay of the
final hadrons (primarily the neutral pion). 
For this reason there has been only one experimental measurement
of photon-photon correlations in heavy ion collisions
obtained at low incident
energy and low photon momenta ($K_T \le$ 20 MeV/$c$) \cite{TAPS-2g}.
In this letter we present first measurements of direct photon 
correlations in central ultra-relativistic heavy-ion collisions.

A detailed description of the layout of the CERN experiment WA98 
can be found in
\cite{WA98-dir}. Here we briefly discuss those subsystems used in the
present analysis. The WA98 photon spectrometer, 
comprising the LEad-glass photon 
Detector Array (LEDA), was located at a distance of 21.5~m downstream 
from the $^{208}$Pb target and provided partial azimuthal coverage over
the rapidity interval $2.35 < y < 2.95$. 
Further downstream, the total transverse 
energy was measured in the MIRAC calorimeter.
The total transverse energy measured in MIRAC was used for 
offline centrality selection.
The analysis presented here was performed
on the 10\% most central $^{208}$Pb+$^{208}$Pb collisions with a total sample 
of $5.8 \times 10^6$ events collected during runs in 1995 and 1996.
No significant correlation signal was observed in the 20\% most peripheral
collision data sample of  $3.9 \times 10^6$ events.

In order to reject most of the hadron background, all showers reconstructed 
in the LEDA spectrometer were required to have a deposited energy of
greater than 750 MeV, well above the minimum ionizing peak energy of 
550 MeV. Hadron showers could be further rejected by the requirement
that the shower have a narrow width, consistent with an electromagnetic
shower in LEDA~\cite{WA98-dir}. In addition, during the 1996 
run period the LEDA charged particle veto was operational and 
provided a shower sample of $>98\%$ photon purity.

\begin{figure}[ht]
  \includegraphics[width=\columnwidth,height=0.85\columnwidth]{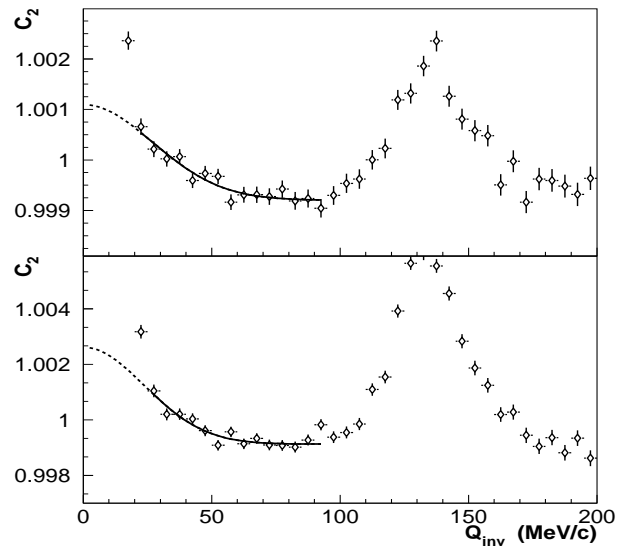} 
\caption{The two-photon correlation function for all showers with 
$L_{min}>20$ cm and average photon momenta
$100 < K_T < 200$  MeV/$c$ (top) and $200 < K_T < 300$ MeV/$c$ (bottom) 
fitted with Eq.~\ref{param}.
The solid line shows the fit result in the fit region used
(excluding the $\pi^0$ peak at $Q_{inv}\approx m_{\pi^0}$) 
and the dotted line shows the extrapolation 
into the low $Q_{inv}$ region where backgrounds are large.
\label{fig:fits}}
\end{figure}

The two-photon correlation function was calculated for each bin of the 
photon average transverse momentum, 
 $K_T=|\vec{p}_{T_1}+\vec{p}_{T_2}|/2$ as the ratio of
the distribution of photon pair invariant relative momenta, $Q_{inv}$, 
where both photons were taken from the same event, to the same
distribution but with the photons of the pair taken from different events.
Sample correlation functions are shown in Fig.~\ref{fig:fits}.
The ratio was normalized to have an equal number of
pairs in the numerator and denominator. 
The correlation function has been fit with a Gaussian parameterisation
\begin{equation}\label{param}
C_2(Q_{inv}) = A \cdot 
[1+\lambda \exp\left ( -R_{inv}^2\cdot Q_{inv}^2 \right ) ]  
\end{equation}
with nomalization $A$, correlation strength $\lambda$, and radius parameter 
$R_{inv}$.

There are a large number of effects which may give rise to small $Q_{inv}$ 
correlations and mimic a direct photon pair
Bose-Einstein correlation. These include: 1) single hadron or photon 
showers that
are split into nearby clusters, 2) photon conversions,  3) HBT 
correlations from charged pions, or other hadrons, mis-identified as photons, 
4) residual photon correlations from $\pi^0$ HBT correlations, 5) radiative
decays of heavier resonances and 6) collective flow. 

Apparatus or analysis effects which may result in the splitting of a single
shower into multiple clusters, or the merging of nearby showers into a single
cluster may be investigated by studying the dependence of the correlation
function on the relative distance $L$ between the showers on the LEDA detector
surface.  These effects are expected to 
contribute strongly at small $L$ and so can be
suppressed effectively by a distance cut. Such a minimum cut on $L$
introduces a lower cutoff in $Q_{inv}$~\cite{cut}

\begin{figure}[ht]
  \includegraphics[width=\columnwidth,height=0.85\columnwidth]{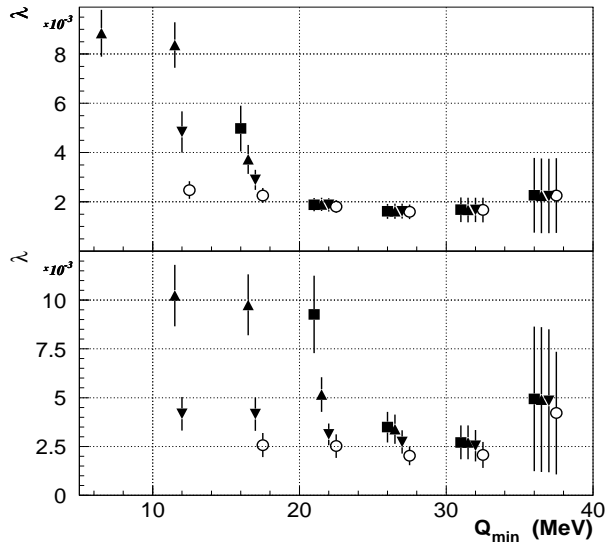}
\caption{Comparison of $\lambda$ parameter fit results for different
fit regions for
$100 < K_T < 200$ MeV/$c$ (top) and
$200 < K_T < 300$ MeV/$c$ (bottom), calculated for narrow showers 
with different cuts on the minimum shower separation distance: $L_{min}$=
20 cm - $\blacksquare$,
25 cm - $\blacktriangle$, 30 cm - $\blacktriangledown$, 
35 cm - $\bigcirc$ (same $Q_{min}$ for each).
\label{fig:ldep}}
\end{figure}

The dependence of the correlation strength parameter $\lambda$ on the minimum
distance cut $L_{min}$ and the minimum invariant momentum $Q_{min}$ used to
define the fit region is shown in Fig.~\ref{fig:ldep} for two $K_T$ regions
for narrow showers.
The different symbols correspond to minimum distance cuts of $L_{min} =
20,25,30,$ and 35 cm (note that a single LEDA module is 4 cm in width).  
The
results demonstrate that the extracted fit parameters vary strongly with
$L_{min}$ when the low $Q_{inv}$ region is included in the fit, a result
attributed to apparatus effects and conversion background, but that stable
results are obtained with a sufficiently large minimum separation 
distance cut, or by
restricting the $Q_{inv}$ fit region. 
When no charged veto or narrow shape cuts are applied to the showers, 
stable results are also obtained, but with larger minimum distance 
cut (or $Q_{min}$) required, consistent with the larger expected backgrounds.

\begin{figure}[ht]
  \includegraphics[width=\columnwidth]{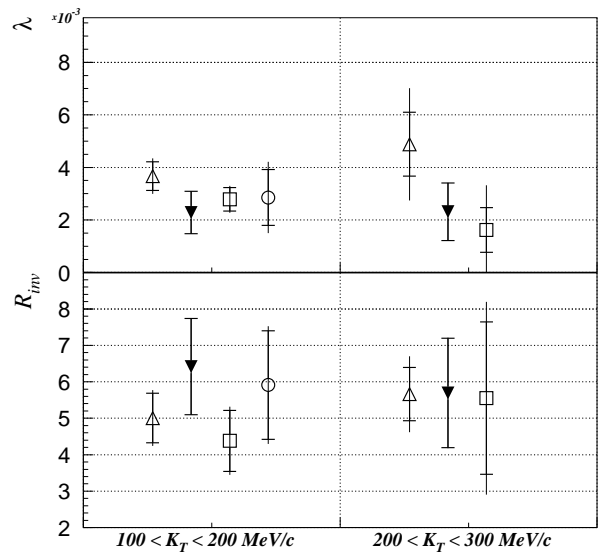}
\caption{Comparison of parameters of correlation functions 
with different particle identification
criteria:  $\triangle$~-~all clusters, $\blacktriangledown$ - 
narrow electromagnetic, $\square$~-~all neutral, $\bigcirc$ - narrow neutral
electromagnetic (no significant result for high $K_T$). \label{fig:pid}}
\end{figure}

As mentioned above, the observed correlations could be caused by residual
correlations of charged pions, neutrons, or conversion electrons misidentified
as photons. 
To investigate possible contributions
from non-photon contamination the correlation functions were constructed with
four different identification criteria applied to the showers reconstructed in
LEDA.  These criteria have somewhat different photon efficiencies, which should
not affect the photon correlation, but more importantly have very different
levels of non-photon contamination which should only affect $\lambda$ 
if the contamination forms uncorrelated background. 
The charged hadron contamination decreases
from $37\%$ and $22\%$ to $16\%$ and $4\%$, respectively for the two $K_T$
bins, after applying the narrow 
electromagnetic shower shape condition~\cite{WA98-dir}
and is negligible after application of the charged veto condition. \
The correlation parameters
extracted from these four types of correlation functions, corrected for 
contamination, are shown in Fig.~\ref{fig:pid}. The consistency
of the parameters extracted with the different identification criteria
indicates that the non-photon contribution to the observed correlation is not
significant.

The estimated systematic errors on the correlation fit parameters are
summarized in Table~\ref{tab:syst}. Besides the uncertainties associated with
the apparatus and fit range discussed with respect to Fig.~\ref{fig:ldep}, and
non-photon contamination (Fig.~\ref{fig:pid}), the dependence on the fit
function has been investigated.  In addition to the Gaussian form of
Eq.~\ref{param}, an exponential form and a Student's t distribution form have
also been used to fit the correlation functions. The Student's t distribution
provides a good fit to the correlation for parameter $n\ge 2$.  The variation
of the fit results with $n$ and comparison with the Gaussian fit results has
been used to estimate the fit function error. Reasonable fit functions leading
to significantly smaller values of $\lambda$ could not be found.
The exponential form gives
larger $\lambda$ values but also exhibits a strong dependence of
the fit parameters on fit range, indicating that it does not reproduce the
shape of the observed correlations well.  Nevertheless, 
if the exponential shape is considered with the limited fit range of $Q_{inv}$,
the upper error bounds on $\lambda$ should be increased, allowing an even 
stronger direct photon signal.

Finally,  correlations might exist in the
background decay photons, e.g. correlations due to collective flow,
Bose-Einstein correlations of $\pi^0$'s, or from decays of heavier resonances.
Monte Carlo simulations have been performed to estimate each of these
effects. For these simulations, the transverse momentum and rapidity
distributions of the $\pi^0$'s were taken from measurement~\cite{WA98-dir}. The
effect of flow was investigated by introducing an elliptic flow pattern with a
magnitude equal to that measured for charged pions \cite{pi-flow}.  Similarly,
the effect of $\pi^0$ Bose-Einstein correlations was introduced with the same
parameters as measured for charged pions~\cite{pich-BE}. Finally, residual
correlations due to decays of heavier resonances, were estimated by including
all resonances having high yield and large branching ratios for electromagnetic
decay: $K^0_S$, $K^0_L$, $\eta$, and $\omega$. The heavy resonances were
included based on experimental spectra where available and thermodynamic
extrapolations otherwise. In all simulations, the acceptance, identification
cuts, and energy and position resolution of LEDA were applied. The simulations
indicated that elliptic flow results in the appearance of a small
slope to the correlation function
on the order of $5\cdot 10^{-6}$ (MeV/c)$^{-1}$;
Bose-Einstein correlations of $\pi^0$'s lead to a specific step-like
correlation function, in agreement with analytical calculations of
\cite{Peressounko}; and residual correlations due to decays of heavier
resonances are found to be 
negligible.  We have checked for flow effects in 
the data sample with highest statistics, fitting with a parameterization with
an additional parameter for the slope, and found a slope parameter
consistent with the
simulations. However, limited statistics did not allow to extend this approach
to all data, and so the final values of the correlation strength and
radii have been corrected for the effect of elliptic flow.

Averaging over the different PID criteria, we obtain the following correlation parameters:
\begin{eqnarray}
\lambda^I&=&0.0028 \pm 0.0004 ({\rm stat.}) \pm 0.0006 ({\rm syst.})  \nonumber \\
R_{inv}^I&=&5.4 \pm 0.8 ({\rm stat.}) \pm 0.9 ({\rm syst.}) {\rm fm}, \nonumber \\
\lambda^{II}&=&0.0029 \pm 0.0007 ({\rm stat.}) \pm 0.0016 ({\rm syst.})  \nonumber \\
R_{inv}^{II}&=&5.8 \pm 0.8 ({\rm stat.}) \pm 1.2 ({\rm syst.}) {\rm fm} \nonumber
\end{eqnarray}
for regions I) $100<K_T<200$~MeV/$c$, and II) $200<K_T<300$~MeV/$c$, respectively. 

\begin{table}
\caption{Summary of systematic errors on the photon-photon correlation parameters.}
\label{tab:syst}
\begin{tabular}{|l|cc|cc|}\hline
                 & \multicolumn{2}{c|}{$100< K_T<200$} &  \multicolumn{2}{c|}{$200< K_T<300$} \\
Source           & $\lambda$ (\%) & $R_{inv}$ (\%)  & $\lambda$ (\%)  & $R_{inv}$ (\%)  \\\hline
Apparatus & 7 & 5 & 16 & 6 \\ 
Contamination & 17 & 14 & 42  & 14 \\ 
Fit Function & 5 & 5 & 18 & 6 \\ 
Fit Range & 8 & 5 & 26 & 10 \\ 
$Q_{inv}$ Slope (Flow)    & 2 & 3 & 12 & 8 \\ \hline
Total Syst. Error (\%) & 21 & 17 & 56 & 21  \\\hline\hline
$ N^{Total}_\gamma$ Total Error (\%)& 12 & - & 4 & -\\\hline
\end{tabular}
\end{table}

The direct photon invariant radii can be compared to measurements of invariant
interferometric radii of $\pi^-$
for the same centrality selection and
$K_T$ region~\cite{WA98-hm}: $R_{inv}=7.11 \pm 0.22, 6.91 \pm 0.32,$ and $6.65 \pm
0.30$ fm at $ K_T = 125, 175,$ and $285$ MeV/$c$, respectively. The similarity
of the interferometric radii of direct photons and pions suggests that the
direct photons of this $K_T$ region are emitted in the late, hadron
gas, stage of the collision.

\begin{figure}[h!t]
  \includegraphics[width=\columnwidth,height=0.85\columnwidth]{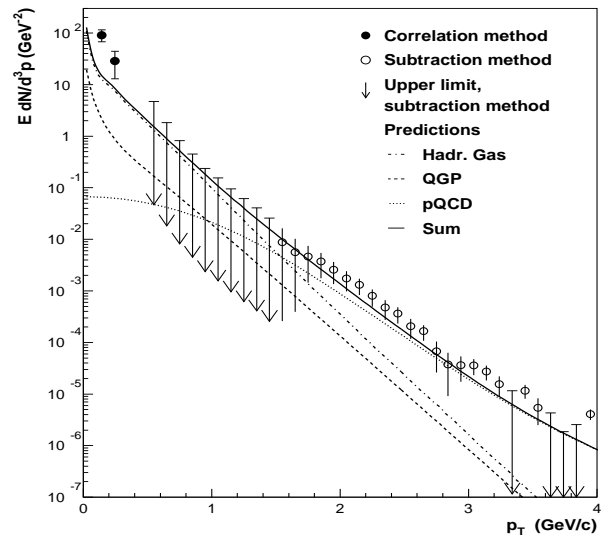}
\caption{Yield of direct photons extracted from the strength of the two-photon 
correlation (closed circles) and
by the statistical subtraction 
method (open circles, or arrows indicating upper limits)~\cite{WA98-dir}. 
Total statistical plus systematical errors are shown. 
The model calculations are described in the 
text.\label{fig:yield}}
\end{figure}

Under the assumption of a fully chaotic photon source, the direct photon yield
$N_\gamma^{Direct}$ is related to the correlation strength $\lambda$ and the
total inclusive photon yield $N_\gamma^{Total}$ as~\cite{Peressounko}
$$
N_\gamma^{Direct}/N_\gamma^{Total}=\sqrt{2\lambda}
$$ 
The low $p_T$ direct photon yield has been extracted using this expression and
is presented in Fig.~\ref{fig:yield} (assuming $p_T = \langle K_T \rangle$).  
The previously published direct photon
yield at high transverse momenta obtained with the subtraction
method~\cite{WA98-dir} is also shown.  The measured direct photon results are
compared with recent fireball model predictions~\cite{Rapp}.  The
calculated contributions to the total yield from the Quark Gluon Plasma
and hadronic stages of the collision are shown. 
It is seen that the contribution from the
hadronic gas phase dominates the direct photon yield at small $p_T$, with
predicted yields below the experimental data.

In summary, two-photon correlation functions have been measured for the first
time in central Pb+Pb collision at 158 A GeV.  The observed correlations are
attributed to Bose-Einstein correlations of directly radiated photons.  
An invariant radius of about 6 fm is
extracted, comparable to that extracted for pions of similar momenta, 
and the correlation strength parameter was used to extract the yield of
direct photons at $p_T<300$ MeV/$c$. The extracted yield exceeds
theoretical expectations which attribute the dominant contribution in this
$p_T$ region to the hadronic phase.

We wish to acknowledge useful discussions with R.~Rapp, C.~Gale, and
D.~Srivastava.
This work was supported jointly by
the German BMBF and DFG, the U.S. DOE, the Swedish NFR and FRN, the
Dutch Stichting FOM, the Polish KBN under Contract No.
621/E-78/SPUB-M/CERN/P-03/DZ211/, the Grant Agency of the Czech
Republic under contract No. 202/95/0217,
the Department of Atomic Energy, the Department of Science and
Technology, the Council of Scientific and Industrial Research and the
University Grants Commission of the Government of India,
the Indo-FRG Exchange Program, the PPE division of CERN, the Swiss
National Fund, the INTAS under Contract INTAS-97-0158, ORISE,
Grant-in-Aid for Scientific Research (Specially Promoted Research \&
International Scientific Research) of the Ministry of Education,
Science and Culture, the University of Tsukuba Special Research
Projects, and the JSPS Research Fellowships for Young Scientists.
ORNL is managed by UT-Battelle, LLC, for the U.S. Department of Energy
under contract DE-AC05-00OR22725.  The MIT group has been supported by
the US Dept.\ of Energy under the cooperative agreement
DE-FC02-94ER40818.

$^{*}$ Deceased.


\begin{thebibliography}{99}


\bibitem{QM-rew} See e.g.\ recent Quark Matter conferences: 
  Nucl.Phys. {\bf A698} (2002),
  J. Phys. {\bf G 27} (2001) 255.
\bibitem{Srivastava}D.K.~Srivastava and J.I.~Kapusta, Phys. Lett. {\bf B 307} (1993)
1; Phys. Rev. {\bf C 48} (1993) 1335; Phys. Rev. {\bf C 50} (1994) 505.
\bibitem{Timerman}A.~Timmermann, M.~Plumer, L.~Razumov and  R.M.~Weiner, Phys.Rev. {\bf C
50} (1994) 3060.
\bibitem{Ornik}U.~Ornik et al., hep-ph/9509367.

\bibitem{Peressounko}D.~Peressounko, Phys. Rev. {\bf C 67} (2003) 014905.

\bibitem{TAPS-2g} M.~Marques et al., Phys. Rep. 284 (1997) 91;
Phys. Rev. Lett. {\bf 73} (1994) 34. Phys. Lett. {\bf B 349} (1995)~30.

\bibitem{WA98-dir}M.M.~Aggarwal et al., Phys. Rev. Lett. {\bf 85} (2000) 3595; 
M.M. Aggarwal et al., nucl-ex/0006007.

\bibitem{cut} Since $Q_{inv}$ is proportional to 
$K_T$, the distance cut implies that the low $Q_{inv}$ Bose-Einstein 
enhancement of the correlation is observable here only 
at low $K_T$. 

\bibitem{pi-flow}H.~Appelshauser et al., Phys.Rev.Lett. {\bf 80} (1998) 4136.

\bibitem{pich-BE}K.~Kadija et al., Nucl. Phys. {\bf A610} (1996) 248c.

\bibitem{WA98-hm} Unpublished WA98 result, analysis follows that described in
L.~Rosselet et al., Nucl. Phys. {\bf A610} (1996) 256c and 
M.M.~Aggarwal et al., Phys. Rev. {\bf C 67} (2003) 014906.

\bibitem{Rapp} S.~Turbide, R.~Rapp, and C.~Gale, hep-ph/0308085.

\end{thebibliography}
\end{document}